\documentclass[conference]{IEEEtran}
\usepackage[T1]{fontenc}
\usepackage[utf8]{inputenc}
\usepackage[binary-units=true]{siunitx}
\DeclareSIUnit{\bits}{bits}
\usepackage{amssymb}
\usepackage[pdftex]{graphicx}
\usepackage{subcaption} 
\usepackage{amsmath}
\usepackage{amsthm}
\usepackage{mathtools}
\usepackage{array}
\usepackage[noadjust]{cite}
\usepackage{xcolor}
\usepackage{tikz}
\usepackage[bookmarks=false]{hyperref}
\usepackage[acronym]{glossaries}
\usepackage{hhline}
\usepackage{multirow}
\usepackage{adjustbox}
\usepackage{pgfplots}
\DeclareUnicodeCharacter{2212}{−}
\usepgfplotslibrary{groupplots,dateplot}
\usetikzlibrary{patterns,shapes.arrows}
\pgfplotsset{compat=newest}

\usepackage{amsfonts} 
\DeclareMathSymbol{\shortminus}{\mathbin}{AMSa}{"39}

\DeclareMathAlphabet{\mymathbb}{U}{bbold}{m}{n}

\usepackage[font=footnotesize]{subcaption}

\newacronym{3GPP}{3GPP}{3rd Generation Partnership Project}
\newacronym{ACM}{ACM}{adaptive coding and modulation}
\newacronym{ADC}{ADC}{analog-to-digital conversion}
\newacronym{AGC}{AGC}{automatic gain control}
\newacronym{AWGN}{AWGN}{additive white Gaussian noise}
\newacronym{BER}{BER}{bit error rate}
\newacronym{BS}{BS}{base station}
\newacronym{BLER}{BLER}{block error rate}
\newacronym{BCE}{BCE}{binary cross-entropy}
\newacronym{BMD}{BMD}{bit-metric decoding}
\newacronym{BP}{BP}{backpropagation}
\newacronym{BPTT}{BPTT}{backpropagation through time}
\newacronym{CE}{CE}{cross-entropy}
\newacronym{CFO}{CFO}{carrier frequency offset}
\newacronym{CNN}{CNN}{convolutional neural network}
\newacronym{CSI}{CSI}{channel state information}
\newacronym{DAC}{DAC}{digital-to-analog conversion}
\newacronym{DL}{DL}{deep learning}
\newacronym{DFT}{DFT}{discrete Fourier transform}
\newacronym{ELU}{ELU}{exponential linear unit}
\newacronym{FFT}{FFT}{fast Fourier transform}
\newacronym{GAN}{GAN}{generative adversarial network}
\newacronym{GRU}{GRU}{gated recurrent unit}
\newacronym{iid}{i.i.d.\@}{independent and identically distributed}
\newacronym{IFFT}{IFFT}{inverse fast Fourier transform}
\newacronym{KL}{KL}{Kullback-Leibler}
\newacronym{LLR}{LLR}{log-likelihood ratio}
\newacronym{LSTM}{LSTM}{long short-term memory}
\newacronym{LDPC}{LDPC}{low-density parity-check}
\newacronym{LMMSE}{LMMSE}{linear minimum mean squared error}
\newacronym{MDP}{MDP}{Markov decision process}
\newacronym{ML}{ML}{machine learning}
\newacronym{MLP}{MLP}{multilayer perceptron}
\newacronym{MIMO}{MIMO}{multiple-input multiple-output}
\newacronym{MU-MIMO}{MU-MIMO}{multi-user multiple-input multiple-output}
\newacronym{MU}{MU}{multi-user}
\newacronym{MSE}{MSE}{mean squared error}
\newacronym{NN}{NN}{neural network}
\newacronym{NR}{NR}{new radio}
\newacronym{NLOS}{NLOS}{non-line of sight}
\newacronym{OFDM}{OFDM}{orthogonal frequency-division multiplexing}
\newacronym{pdf}{pdf}{probability density function}
\newacronym{pmf}{pmf}{probability mass function}
\newacronym{QPSK}{QPSK}{quadrature phase-shift keying}
\newacronym{QAM}{QAM}{quadrature amplitude modulation}
\newacronym{PSNR}{PSNR}{Peak Signal to Noise Ratio}
\newacronym{RBF}{RBF}{Rayleigh block-fading}
\newacronym{RB}{RB}{resource block}
\newacronym{RE}{RE}{resource element}
\newacronym{RG}{RG}{resource grid\newacronym{RE}{RE}{resource element}}
\newacronym{ReLU}{ReLU}{rectified linear unit}
\newacronym{RTN}{RTN}{radio transformer network}
\newacronym{RL}{RL}{reinforcement learning}
\newacronym{RNN}{RNN}{recurrent neural network}
\newacronym{SFO}{SFO}{sampling frequency offset}
\newacronym{SER}{SER}{symbol error rate}
\newacronym{SNR}{SNR}{signal-to-noise ratio}
\newacronym{SINR}{SINR}{signal-to-interference-plus-noise ratio}
\newacronym{SGD}{SGD}{stochastic gradient descent}
\newacronym{SISO}{SISO}{single-input single-output}
\newacronym{SIMO}{SIMO}{single-input multiple-output}
\newacronym{SU}{SU}{single-user}
\newacronym{TDD}{TDD}{time-division duplexing}
\newacronym{TR}{TR}{technical report}
\newacronym{UE}{UE}{user equipment}
\newacronym{UMi}{UMi}{urban microcell}
\newacronym{wrt}{w.r.t.\@}{with respect to}

\renewcommand{\vec}[1]{\mathbf{#1}}
\newcommand{\vecs}[1]{\boldsymbol{#1}}

\newcommand{\hv}{\vec{h}}

\newcommand{\nv}{\vec{n}}

\newcommand{\tv}{\vec{t}}

\newcommand{\vv}{\vec{v}}
\newcommand{\wv}{\vec{w}}
\newcommand{\xv}{\vec{x}}
\newcommand{\yv}{\vec{y}}

\newcommand{\thetav}{\vecs{\theta}}

\newcommand{\Dm}{\vec{D}}
\newcommand{\Em}{\vec{E}}

\newcommand{\Hm}{\vec{H}}
\newcommand{\Id}{\vec{I}}

\newcommand{\Tm}{\vec{T}}

\newcommand{\Wm}{\vec{W}}
\newcommand{\Xm}{\vec{X}}
\newcommand{\Ym}{\vec{Y}}

\newcommand{\Sigmam}{\vecs{\Sigma}}

\newcommand{\CC}{\mathbb{C}}

\newcommand{\RR}{\mathbb{R}}

\newcommand{\htp}{^{\mathsf{H}}}
\newcommand{\tp}{^{\mathsf{T}}}

\newcommand{\LB}{\left(}
\newcommand{\RB}{\right)}

\newcommand{\LSB}{\left[}
\newcommand{\RSB}{\right]}

\renewcommand{\log}[1]{\mathop{\mathrm{log}}\LB #1\RB}
\renewcommand{\exp}[1]{\mathop{\mathrm{exp}}\LB #1\RB}

\newcommand{\EE}{{\mathbb{E}}}

\newlength{\dhatheight}

\IEEEoverridecommandlockouts
\begin{document}
\begin{NoHyper}
\title{Machine Learning-enhanced Receive Processing\\for MU-MIMO OFDM Systems}

\author{
\IEEEauthorblockN{Mathieu Goutay\IEEEauthorrefmark{1}\IEEEauthorrefmark{3}, Fayçal Ait Aoudia\IEEEauthorrefmark{1}, Jakob Hoydis\IEEEauthorrefmark{2}\thanks{Work carried out while J. Hoydis was with Nokia Bell Labs.}, and Jean-Marie Gorce\IEEEauthorrefmark{3}}
\IEEEauthorblockA{\IEEEauthorrefmark{1}Nokia Bell Labs, Paris-Saclay, 91620 Nozay, France}
\IEEEauthorblockA{\IEEEauthorrefmark{2}NVIDIA, 06906 Sophia Antipolis, France}
\IEEEauthorblockA{\IEEEauthorrefmark{3}Université de Lyon, INSA Lyon, Inria, CITI,  69100 Villeurbanne, France \\
\{mathieu.goutay, faycal.ait\_aoudia\}@nokia.com, jhoydis@nvidia.com, jean-marie.gorce@insa-lyon.fr
}}

\maketitle

\begin{abstract}

\Gls{ML} can be used in various ways to improve \gls{MU-MIMO} receive processing.
Typical approaches either augment a single processing step, such as symbol detection, or replace multiple steps jointly by a single \gls{NN}.
These techniques demonstrate promising results but often assume perfect \gls{CSI} or fail to satisfy the interpretability and scalability constraints imposed by practical systems.
In this paper, we propose a new strategy which preserves the benefits of a conventional receiver, but enhances specific parts with \gls{ML} components.
The key idea is to exploit the \gls{OFDM} signal structure to improve both the demapping and the computation of the channel estimation error statistics.
Evaluation results show that the proposed \gls{ML}-enhanced receiver beats practical baselines on all considered scenarios, with significant gains at high speeds.


\end{abstract}

\glsresetall

\section{Introduction} 
\label{sec:introduction}

Future generations of wireless networks will need to handle the growing demand for connectivity.
At the physical layer, \glsdesc{MU}\glsunset{MU} \glsdesc{MIMO}\glsunset{MIMO} (MU-MIMO) is a promising technique to increase the number of users that can be served simultaneously.
Linear methods are often used to decrease the computational complexity of the receive processing, but they achieve poor performance on realistic channels. 
It is therefore crucial to find new solutions that both satisfy the constraints of practical deployments and the performance requirements of future wireless communication systems.

Motivated by the successes of \gls{ML} when applied to the physical layer \cite{9078454}, two \gls{ML}-based approaches have been proposed to improve \gls{MIMO} reception.
The first one consists in augmenting a single processing step of a conventional \gls{MIMO} receiver using a \gls{NN}.
This strategy has been applied to separately improve channel estimation \cite{8752012, mashhadi2020pruning}, equalization \cite{pratik2020remimo, 8646357, MMNet}, and demapping \cite{shental2020machine}.
Although encouraging results have been shown, the proposed solutions usually require perfect \gls{CSI}, remain too complex for practical deployments, or are designed for \gls{SISO} systems only.

The second approach is to replace multiple receive components by a single \gls{NN}.
For example, \cite{korpi2020deeprx} demonstrates strong results by replacing the channel estimation, equalization, and demapping steps by a \gls{CNN} coupled with a so-called transformation layer.
The main advantage is that the \gls{NN} is directly optimized to improve the accuracy of the estimated bits, but the counterpart is that the number of connected users is dictated by the \gls{NN} architecture.


In this paper, we introduce a new strategy which aims to combine the advantages of both approaches while avoiding their shortcomings.
The key idea is to use several \gls{ML} components to enhance specific parts of a conventional \gls{MU}-\gls{MIMO} architecture. 
More precisely, the \gls{OFDM} signal structure is exploited by multiple \glspl{CNN} to improve two receive processing steps.
The first one is the computation of the channel estimation error second order statistics, that the \glspl{CNN} are able to learn during training.
The second one is the demapping, which is carried out by a \gls{CNN} processing the entire \gls{OFDM} time-frequency grid instead of individual resource elements.
All \glspl{CNN} are jointly optimized to maximize the information rate of the transmission \cite{9118963}.

The proposed \gls{ML}-enhanced receiver is benchmarked against two conventional receivers, the second one having perfect \gls{CSI}. 
A 3GPP-compliant channel model was considered, with two pilot configurations and users moving at speeds ranging from $0$ to \SI{130}{\km\per\hour}.
The results indicate that the \gls{ML}-enhanced receiver beats the baseline on every considered scenario and enables gains that increase with the user speeds.

\textbf{Notations :} 
We denote by $\Tm_{a, b} \in \CC^{N_c \times N_d}$ ($\tv_{a, b, c} \in \CC^{N_d}$, $t_{a, b, c, d} \in \CC$) the matrix (vector, scalar) formed by slicing the tensor $\Tm \in \CC^{N_a \times N_b \times N_c \times N_d}$ along the first two (three, four) dimensions.
The notation $ \Tm^{(k)}$ indicates that the quantity at hand is only considered for the $k^{\text{th}}$ user, and $\vv_{\shortminus a}$ is the vector $\vv$ without its $a^{\text{th}}$ element.
$\Id_N$ is the $N \times N$ identity matrix and $\mymathbb{1}_{N \times M}$ is the $N\times M$ matrix with all elements set to $1$.

\section{Channel Model} 
\label{sec:channel_model}

We consider a \gls{MU}-\gls{MIMO} system where $N_k$ single-antenna users transmit \gls{OFDM} signals to a \gls{BS} comprising $N_m$ antennas.
The signals are transmitted over $N_t$ \gls{OFDM} symbols and $N_f$ subcarriers, and the overall time-frequency grid is called \gls{RG} and is illustrated in Fig.~\ref{fig:resource_grid}.
A \gls{RE} refers to one cell of the \gls{RG}, and a group of $12$ adjacent subcarriers is called a resource block.
The channel corresponding to one \gls{RE} $(f,t)$ is denoted by $\Hm_{f,t}\in \CC^{N_m \times N_k}$, and is a slice of the 4-dimensional tensor $\Hm \in \CC^{N_f \times N_t \times N_m \times N_k}$ containing the channel coefficients of the entire \gls{RG}.
Following similar notations, the signal vectors sent by the users and received by the \gls{BS} are respectively denoted by $\xv_{f,t}\in \CC^{N_k}$ and by $\yv_{f,t}\in \CC^{N_m}$.
$M$ denotes the modulation order of the transmission.
The channel transfer function on the \gls{RE} $(f,t)$ is expressed as
\begin{equation}
\label{eq:transfert_function}
\yv_{f,t} = \Hm_{f,t} \xv_{f,t} + \nv_{f,t}
\end{equation}
where $\nv_{f,t} \thicksim \mathcal{C}\mathcal{N}(\mathbf{0}, \sigma^2 \Id_{N_m})$ is the noise vector with a power $\sigma^2$ assumed equal for all users and all \glspl{RE}.
It is assumed that all users have perfect power control such that the mean energy corresponding to a single user and receiving antenna is equal to one, i.e., $\EE\LSB | h_{f,t,k,m}|^2 \RSB =1$.
The \gls{SNR} of the transmission is defined as $\text{SNR} =  10 \log{\frac{1}{\sigma^2}} \, [\si{dB}]$.

\begin{figure}[t]
	
  	\begin{subfigure}{0.2\textwidth}
  	\center
    	\includegraphics[height=110pt]{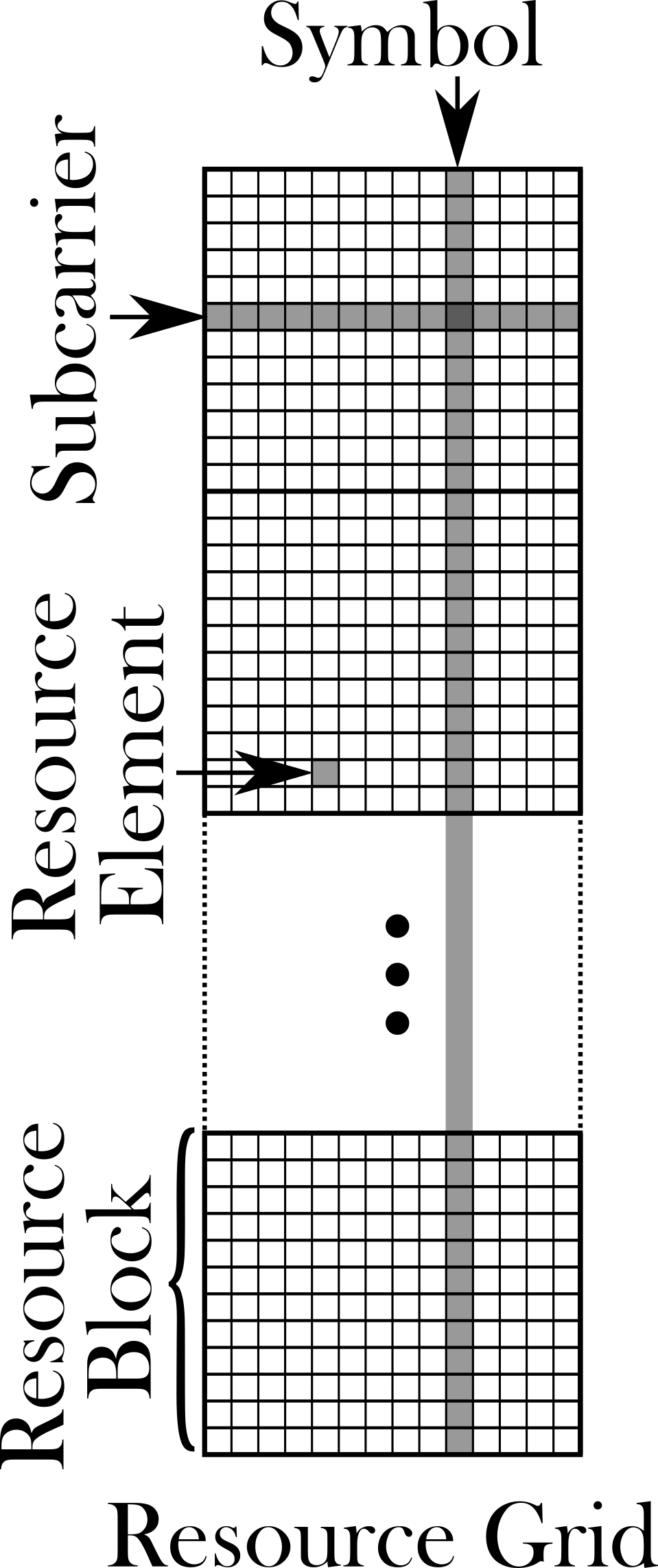} 
  	\caption{Channel nomenclature.}
  	\label{fig:resource_grid}
	\end{subfigure}%
\hspace{0pt}
  	\begin{subfigure}{0.2\textwidth}
  	\center
  	\includegraphics[height=110pt]{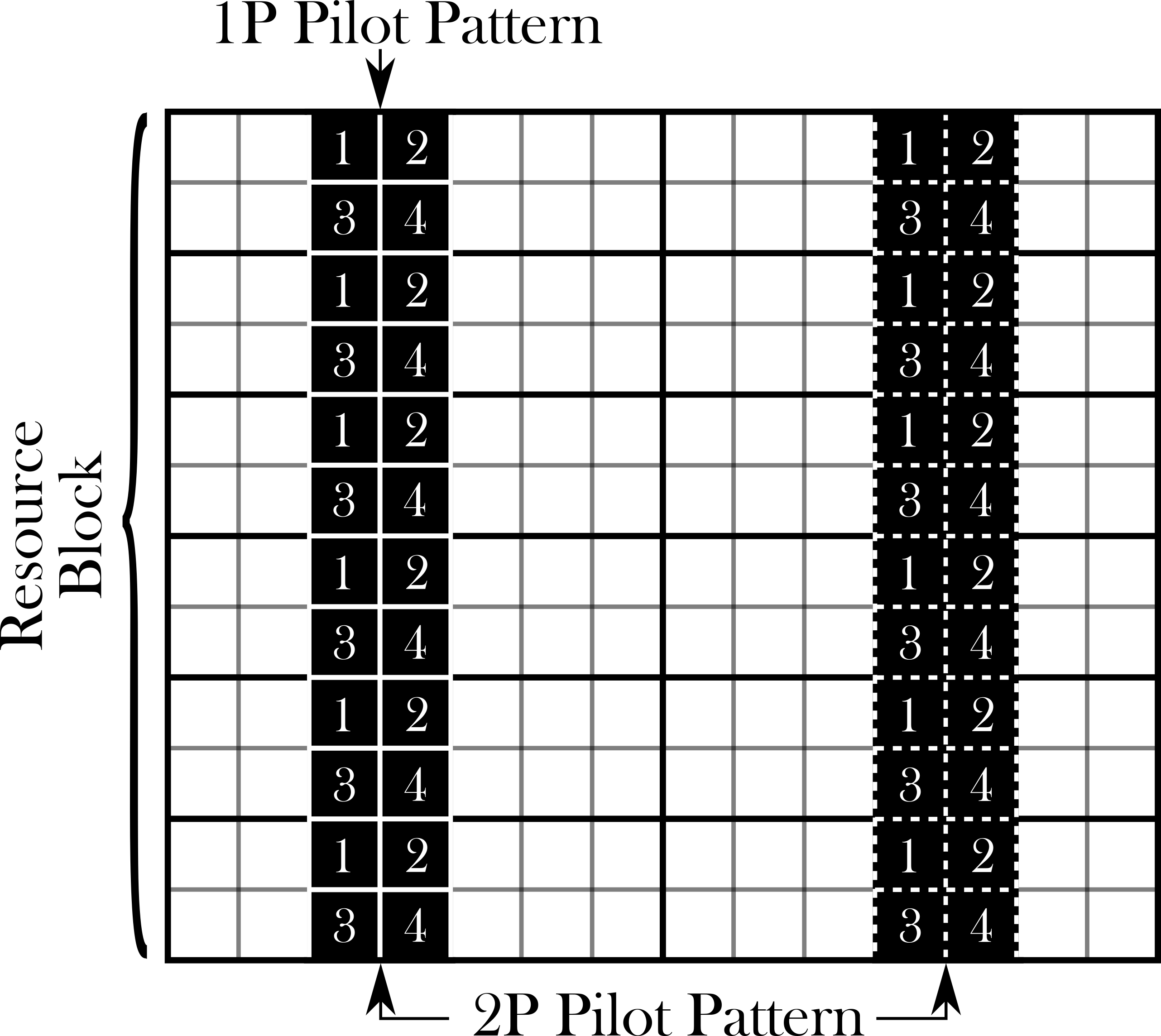}
  	\caption{1P and 2P pilot patterns.}
  	\label{fig:pilot_pattern}
	\end{subfigure}%

\caption{Pilots are arranged on the \gls{RG} following two distinct patterns, where the numbers represent different transmitters.}
\label{fig:channel_model}
\vspace{-10pt}
\end{figure}

Users both transmit data signals and pilot signals, the latter being assumed equal to one and are used by the \gls{BS} to estimate the channel.
Two pilot patterns are considered in this work, referred to as 1P and 2P, and depicted in Fig.~\ref{fig:pilot_pattern}.
Each user $k$ sends pilots on a set of \glspl{RE} denoted by $\mathcal{P}^{(k)}$, and the number of \gls{OFDM} symbols and subcarriers used to carry such pilots  are respectively denoted by $N_{P_t}$ and $N_{P_f}$.
For example, if only one resource block is considered and the 1P pilot pattern is used, then $\mathcal{P}^{(1)} = \{(1, 3), (3, 3), (5,3), (7, 3), (9, 3), (11,3)\}$, $N_{P_t} = 1$, and $N_{P_f} = 6$ (see Fig.~\ref{fig:pilot_pattern}).
The \glspl{RE} used by one user to transmit pilots are not used by the other users such that the pilots do not suffer from interferences.

\section{Conventional Receiver Architecture} 
\label{sec:conventional_architecture}


\subsection{Channel estimation}
\label{sec:ch_est_1}

The first step performed by a conventional receiver is to estimate the channel.
As the pilots are orthogonal, channel estimation can be carried out separately for each user. 
The channel covariance matrix providing the spatial, temporal, and spectral correlation between all \glspl{RE} carrying pilots is denoted by $\Sigmam \in \CC^{N_{P_f}N_{P_t}N_m \times N_{P_f}N_{P_t}N_m}$.
This covariance matrix can be estimated by gathering a large dataset of received pilot signals, in contrast to the covariance between all \glspl{RE} that is usually not available in practice.
To obtain the channel estimate $\widehat{\Hm}^{(k)}_{\mathcal{P}^{(k)}}$ of user $k$ at \glspl{RE} carrying pilots, a well known estimator is the \gls{LMMSE} filter:
\begin{equation}
\text{vec}\LB\widehat{\Hm}^{(k)}_{\mathcal{P}^{(k)}}\RB = \Sigmam \LB \Sigmam  + \sigma^2 \Id \RB^{-1} \text{vec}\LB\Ym^{(k)}_{\mathcal{P}^{(k)}}\RB
\end{equation}
where $\Ym^{(k)}_{\mathcal{P}^{(k)}} \in \CC^{N_{P_f} \times N_{P_t} \times N_m}$ is the tensor of received pilots for user $k$.
The channel estimates for all \glspl{RE} are computed by first linearly interpolating the estimates from \glspl{RE} carrying pilots in the frequency dimension, and then using for each other \glspl{RE} the estimates computed at their nearest interpolated resource element (NIRE).
It is also possible to leverage temporal linear interpolation when the 2P pilot pattern is used.
The tensor of channel estimates for user $k$ is denoted by $\widehat{\Hm}^{(k)} \in \CC^{N_f \times N_t \times N_m}$, and the tensor of channel estimates for all users $\widehat{\Hm} \in \CC^{N_f \times N_t \times N_k \times N_m}$  is obtained by stacking all $\widehat{\Hm}^{(k)}$.
The tensor of corresponding channel estimation errors is denoted by $\widetilde{\Hm} \in \CC^{N_f \times N_t \times N_k \times N_m}$.

The \emph{spatial} channel estimation error covariance is 
\begin{equation}
\label{eq:E}
\Em_{f,t} \coloneqq \sum_{k=1}^{N_k}\Em_{f,t}^{(k)} = \sum_{k=1}^{N_k}\EE \LSB \tilde{\hv}^{(k)}_{f, t} \tilde{\hv}^{(k)^{\scriptstyle \mathsf{H}}}_{f, t} \RSB \in \CC^{N_m \times N_m}
\end{equation}
for the \gls{RE} $(f,t)$, which is the sum of the spatial covariances $\Em_{f,t}^{(k)}$ computed for each user.
These covariance matrices only reflect the correlations between antennas, and not between different \gls{OFDM} symbols or subcarriers.
They are estimated on \glspl{RE} carrying pilots for each user independently, and shared by the groups of \glspl{RE} delimited by thick lines in Fig.~\ref{fig:pilot_pattern} following a nearest-pilot approximation.
The estimated error covariance for a \gls{RE} $(f,t)$ is denoted by  $\widehat{\Em}_{f,t} \in \CC^{N_m \times N_m}$.
The estimation procedure is shown in Fig.~\ref{fig:ch_est_tradi} and is further detailed in \cite{goutay2020machine}.

\begin{figure}[t]
\vspace{10pt}
    \centering
    \includegraphics[width=0.5\textwidth]{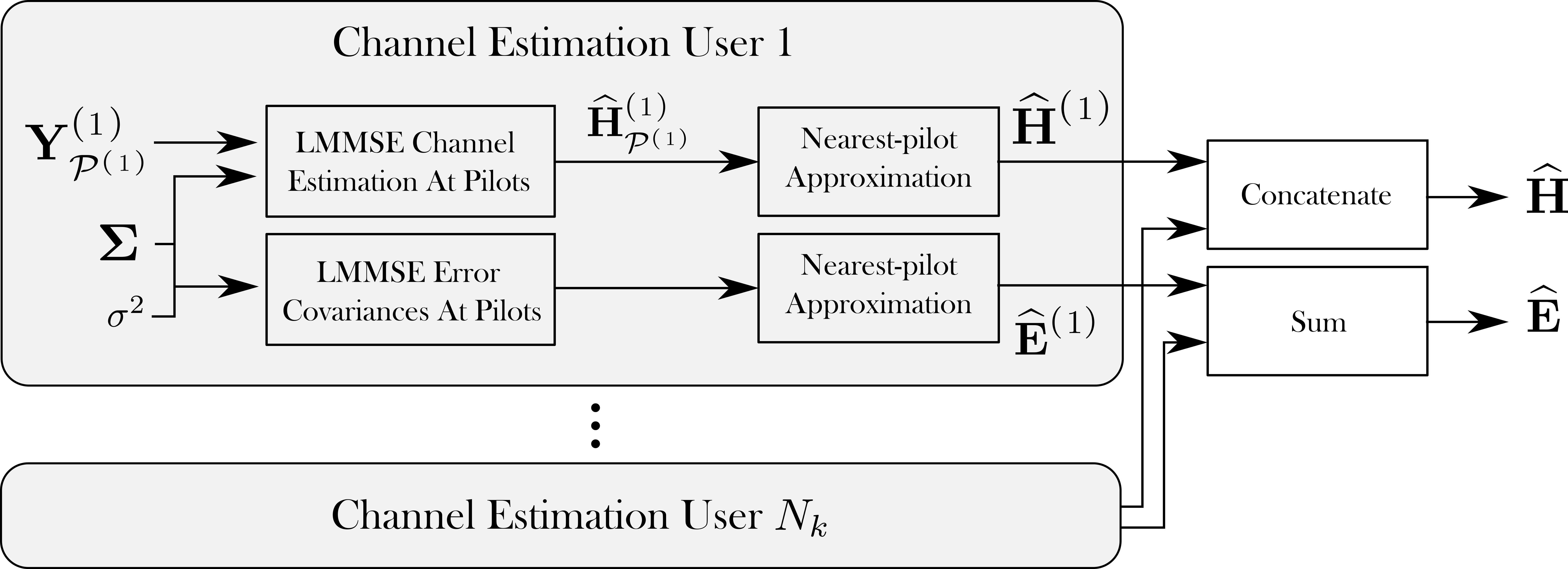}
    \caption{Conventional channel estimation.}
    \label{fig:ch_est_tradi}
    \vspace{-15pt}
\end{figure}

\subsection{Equalization and demapping}

Once the channel has been estimated, the equalization step aims to convert the \gls{MIMO} channel on each \gls{RE} into $N_k$ independent \gls{AWGN} channels.
Linear equalizers such as \gls{LMMSE} are widely used because of their reasonable complexity.
However, the standard \gls{LMMSE} operator  requires the computation of a matrix inversion per \gls{RE}, which can be too computationally expensive in large systems.
Therefore, we leverage a \emph{grouped}-\gls{LMMSE} equalizer, where a single operator $\Wm_{f,t}\in \CC^{N_k \times N_m}$ is computed and applied to a group of \glspl{RE} spanning $\{F_b,\dots, F_e\}\times\{T_b,\dots,T_e\}$, as delimited in Fig.~\ref{fig:pilot_pattern}.
Assuming perfect knowledge of  $\Em$, such operator is computed as (see \cite{goutay2020machine} for a derivation)
\begin{align}
\label{eq:W}
\footnotesize
\begin{split}
\Wm_{f,t} = & \LB \sum_{f'=F_b}^{F_e} \sum_{t'=T_b}^{T_e}  \widehat{\Hm}_{f', t'}\htp \RB \\
&\LB \sum_{f'=F_b}^{F_e} \sum_{t'=T_b}^{T_e}  \widehat{\Hm}_{f', t'} \widehat{\Hm}_{f',t'}\htp + \Em_{f',t'} + \sigma^2 \Id_{N_m}  \RB^{-1}.
\end{split}
\end{align}
In order to obtain $N_k$ \gls{AWGN} channels, the equalized symbols $\hat{\xv}_{f,t}$ needs to be scaled such that $\hat{x}_{f,t,k} = x_{f,t,k} + z_{f,t,k}$, where $z_{f,t,k} \thicksim \mathcal{C}\mathcal{N}(0, \rho_{f, t, k}^2)$ includes the interference and  noise of user $k$. \footnote{This is not true in general as the interference and channel estimation errors are not Gaussian distributed.}
The corresponding scaling matrix 
\begin{equation}
\Dm_{f,t} = \LB  \LB \Wm_{f,t} \hat{\Hm}_{f, t} \RB \odot \Id_{N_k} \RB^{-1}
\end{equation}
is thus applied to the equalized signals, i.e.,
\begin{equation}
\hat{\xv}_{f, t} = \Dm_{f,t} \Wm_{f,t} \yv_{f,t}
\end{equation}
and the post-equalization noise variance $\rho_{f, t, k}^2$ is given by
\small
\begin{equation}
\label{eq:eq_snr_ul}
\rho_{f, t, k}^2  = \frac{\wv_{f, t, k}\htp \LB\widehat{\Hm}_{f, t, \shortminus k} \widehat{\Hm}_{f, t, \shortminus k}\htp + \Em_{f,t} + \sigma^2 \Id_{N_m} \RB \wv_{f, t, k}}{\wv_{f, t, k}\htp \widehat{\hv}_{f, t, k} \widehat{\hv}_{f, t, k}\htp \wv_{f, t, k}}.
\end{equation}
\normalsize
\Glspl{LLR} are obtained using a standard \gls{AWGN} demapper on each user $k$ and \gls{RE} $(f,t)$ independently, assuming a post-equalization noise variance of $\rho_{f, t, k}^2$.
Finally, $\Em_{f,t}$ is often not available and is replaced by its estimate $\hat{\Em}_{f,t}$.

\section{ML-Enhanced receiver architecture}
\label{sec:ml-receiver}

\begin{figure}[t]
    \centering
    \includegraphics[width=0.49\textwidth]{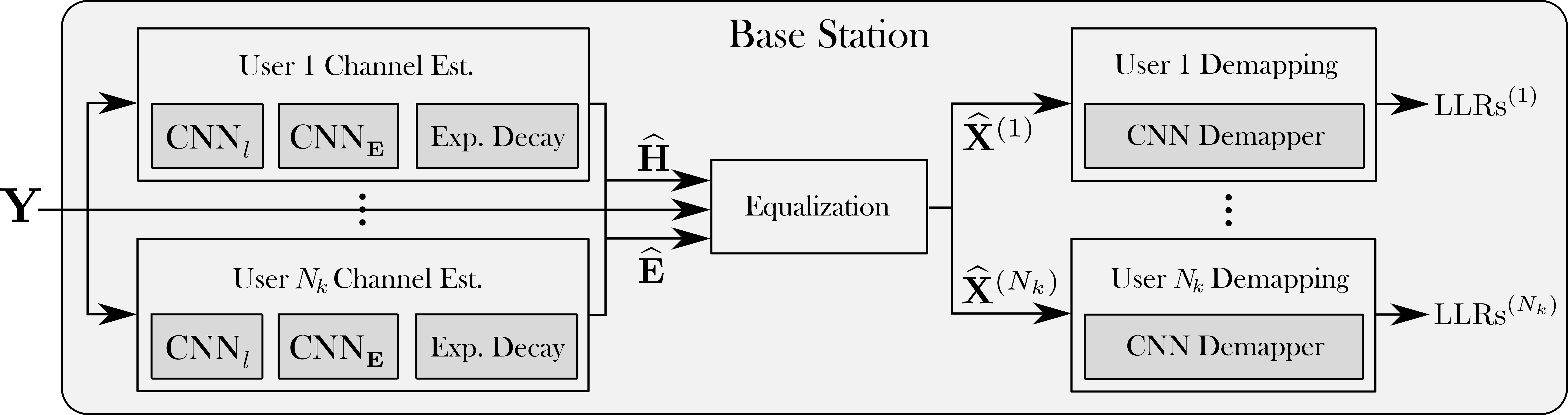}
    \caption{ML-enhanced receiver architecture.}
    \label{fig:ml_receiver}
    \vspace{-10pt}
\end{figure}


\subsection{Receiver training}

The proposed \gls{ML}-enhanced receiver is depicted in Fig.~\ref{fig:ml_receiver}, where the grayed elements represent trainable components.
In contrast to many related papers that optimize each receive processing block separately, we chose to train all \gls{ML} components together to optimize the estimated \glspl{LLR}.
This approach is more practical since it does not assume that ground-truth channel measurements are available.
The set of all trainable parameters $\thetav$ is optimized with stochastic gradient descent on the binary cross-entropy loss and using batches of $B_s$ samples:

\vspace{-10pt}
\small
\begin{equation}
\label{eq:loss_mc}
\mathcal{L} = - \frac{1}{B_s} \sum_{s=1}^{B_s} \sum_{k=1}^{N_k} \sum_{(f, t)\in \mathcal{D}} \sum_{m=1}^{M} \text{log}_2 \LB 
\widetilde{P}_{\thetav} \LB b_{f,t,k,m}^{[s]} | \Ym^{[s]} \RB \RB 
\end{equation}
\normalsize
where $\mathcal{D}$ denotes the set of \glspl{RE} carrying data, $b_{f,t,k,m}$ the $m^{\text{th}}$ bit of user $k$ on the \gls{RE} $(f,t)$, and the superscript $ [s] $ refers to the $s^{\text{th}}$ sample in the batch.
$\widetilde{P}_{\thetav} ( b_{f,t,k,m}^{[s]} | \Ym^{[s]} )$ is the estimated posterior probabilities on the bit $b_{f,t,k,m}^{[s]}$ and is obtained by applying the sigmoid function to the corresponding \glspl{LLR}.
As detailed in \cite{9118963}, minimizing this loss is equivalent to maximizing the sum of the achievable rates for all users. 
%


\subsection{ML-enhanced channel estimator}

\begin{figure}[t]
    \centering
    \includegraphics[width=0.5\textwidth]{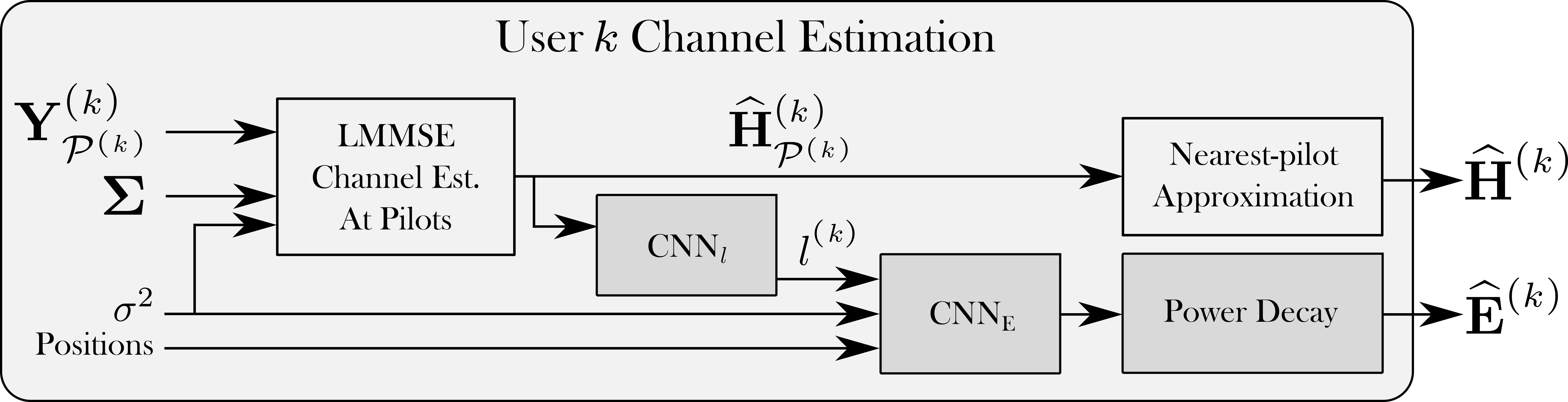}
    \caption{ML-enhanced channel estimation of user $k$.}
    \label{fig:ch_est_ml}
    \vspace{-10pt}
\end{figure}

\begin{figure}[b]
\vspace{-10pt}
\centering
	
  	\begin{subfigure}{0.2\textwidth}
  	\centering
    	\begin{adjustbox}{height=0.7\textwidth} 
\begin{tikzpicture}

\begin{axis}[
colorbar,
colorbar style={ylabel={}},
colormap/viridis,
point meta max=0.1654548958409578,
point meta min=0.0362763414159417,
tick align=outside,
tick pos=left,
x grid style={white!69.0196078431373!black},
xmin=0.5, xmax=16.5,
xtick style={color=black},
y dir=reverse,
y grid style={white!69.0196078431373!black},
ymin=0.5, ymax=16.5,
ytick style={color=black},
width=7cm,height=7cm
]
\addplot graphics [includegraphics cmd=\pgfimage,xmin=0.5, xmax=16.5, ymin=16.5, ymax=0.5] {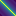};
\end{axis}

\end{tikzpicture}
  	\end{adjustbox}    
  	\caption{Amplitude}
	\end{subfigure}%
\hspace{28pt}
  	\begin{subfigure}{0.2\textwidth}
  	\centering
  	\begin{adjustbox}{height=0.7\textwidth} 
\begin{tikzpicture}

\begin{axis}[
colorbar,
colorbar style={ylabel={}},
colormap={mymap}{[1pt]
  rgb(0pt)=(0.2298057,0.298717966,0.753683153);
  rgb(1pt)=(0.26623388,0.353094838,0.801466763);
  rgb(2pt)=(0.30386891,0.406535296,0.84495867);
  rgb(3pt)=(0.342804478,0.458757618,0.883725899);
  rgb(4pt)=(0.38301334,0.50941904,0.917387822);
  rgb(5pt)=(0.424369608,0.558148092,0.945619588);
  rgb(6pt)=(0.46666708,0.604562568,0.968154911);
  rgb(7pt)=(0.509635204,0.648280772,0.98478814);
  rgb(8pt)=(0.552953156,0.688929332,0.995375608);
  rgb(9pt)=(0.596262162,0.726149107,0.999836203);
  rgb(10pt)=(0.639176211,0.759599947,0.998151185);
  rgb(11pt)=(0.681291281,0.788964712,0.990363227);
  rgb(12pt)=(0.722193294,0.813952739,0.976574709);
  rgb(13pt)=(0.761464949,0.834302879,0.956945269);
  rgb(14pt)=(0.798691636,0.849786142,0.931688648);
  rgb(15pt)=(0.833466556,0.860207984,0.901068838);
  rgb(16pt)=(0.865395197,0.86541021,0.865395561);
  rgb(17pt)=(0.897787179,0.848937047,0.820880546);
  rgb(18pt)=(0.924127593,0.827384882,0.774508472);
  rgb(19pt)=(0.944468518,0.800927443,0.726736146);
  rgb(20pt)=(0.958852946,0.769767752,0.678007945);
  rgb(21pt)=(0.96732803,0.734132809,0.628751763);
  rgb(22pt)=(0.969954137,0.694266682,0.579375448);
  rgb(23pt)=(0.966811177,0.650421156,0.530263762);
  rgb(24pt)=(0.958003065,0.602842431,0.481775914);
  rgb(25pt)=(0.943660866,0.551750968,0.434243684);
  rgb(26pt)=(0.923944917,0.49730856,0.387970225);
  rgb(27pt)=(0.89904617,0.439559467,0.343229596);
  rgb(28pt)=(0.869186849,0.378313092,0.300267182);
  rgb(29pt)=(0.834620542,0.312874446,0.259301199);
  rgb(30pt)=(0.795631745,0.24128379,0.220525627);
  rgb(31pt)=(0.752534934,0.157246067,0.184115123);
  rgb(32pt)=(0.705673158,0.01555616,0.150232812)
},
point meta max=2.44783252080282,
point meta min=-2.19349443912506,
tick align=outside,
tick pos=left,
x grid style={white!69.0196078431373!black},
xmin=0.5, xmax=16.5,
xtick style={color=black},
y dir=reverse,
y grid style={white!69.0196078431373!black},
ymin=0.5, ymax=16.5,
ytick style={color=black},
width=7cm,height=7cm
]
\addplot graphics [includegraphics cmd=\pgfimage,xmin=0.5, xmax=16.5, ymin=16.5, ymax=0.5] {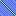};
\end{axis}

\end{tikzpicture}
  	\end{adjustbox} 
  	\caption{Phase}
	\end{subfigure}%

\caption{Example of amplitude and phase for $\Em_{f,t}^{(k)}$.}
\label{fig:E}
\end{figure}

The NIRE approximation used by the conventional channel estimator of Section~\ref{sec:conventional_architecture} implies that the estimation of the covariances $\Em_{f,t}$ might be inaccurate at \glspl{RE} that are far from pilots.
This in turn causes the computation of $\Wm_{f,t}$ in \eqref{eq:W} and of $\rho^2_{f,t,k}$ in \eqref{eq:eq_snr_ul} to be erroneous.
In the following, we present a suite of \gls{ML} components, represented in grey in Fig.~\ref{fig:ch_est_ml}, that aim to better predict these channel estimation error covariances.

Fig.~\ref{fig:E} shows an example of the amplitudes and phases of an error covariance matrix $\Em^{(k)}_{f,t}$.
Predicting the complex coefficient of $\Em_{f,t}$ for all \glspl{RE} would be of prohibitive complexity for a naive \gls{NN}.
To tackle this problem, we approximate every element $(x,y)$ of $\Em_{f,t}$ with a complex power decay model:
\begin{equation}
\label{eq:exp_decay}
\hat{e}^{(k)}_{f,t,x,y}= \alpha_{f,t} \beta_{f,t}^{|y-x|} \exp{j \gamma (y-x)}
\end{equation}
where $\alpha_{f,t}$ and $\beta_{f,t}$ are parameters that control the scale and the decay of the model, and are different for every \gls{RE} $(f,t)$.
We observed that the phase offset $\gamma$ was close to $\pi$ and constant over the \gls{RG}, and is therefore implemented as a single trainable parameter that will be optimized according to \eqref{eq:loss_mc}.

To estimate  $\alpha_{f,t}$ and $\beta_{f,t}$ for all \glspl{RE}, we propose to use a \gls{CNN}, denoted by $\text{CNN}_{\Em}$.
This \gls{CNN} has an output of dimension of $N_f \times N_t \times 2$, corresponding to the two parameters being predicted on the entire \gls{RG}, and has four inputs of dimension $N_f \times N_f$, for a total input dimension of $N_f \times N_f \times 4$.
\glspl{CNN} are known to be translation invariant, however the predictions $\alpha_{f,t}$ and $\beta_{f,t}$ depend on the position $(f,t)$ in the \gls{RG}.
Therefore, vertical and horizontal positional information are given to $\text{CNN}_{\Em}$ by the first two input matrices, that respectively have all columns equal to $\LSB -\frac{N_f}{2}, \cdots, -1, 1, \cdots, \frac{N_f}{2} \RSB\tp$ and all rows equal to $\LSB \frac{N_t}{2}, \cdots, -1, 1, \cdots, \frac{N_t}{2} \RSB$.
The third input is the \gls{SNR} of the transmission, given as $\text{SNR} \cdot \mymathbb{1}_{N_f \times N_t}$.
The fourth input is designed to provide information about the time and frequency selectivity of the channel, since the NIRE approximation of $\widehat{\Hm}^{(k)}$ is likely to be less precise on fast-varying channels.

This time-variability information is in turn estimated by a second \gls{CNN}, denoted by $\text{CNN}_{l}$.
The goal of this \gls{CNN} is to extract a feature related to the Doppler and delay spread of the transmission from the estimated channel at \glspl{RE} carrying pilots.
This feature is a single scalar denoted by $l^{(k)} \in \RR $ for user $k$, as shown in Fig.~\ref{fig:ch_est_ml}.
$\text{CNN}_{l}$ has inputs of dimension $N_{P_f} \times N_{P_t} \times 2N_m$, corresponding to the stacking of the real and imaginary part of $\widehat{\Hm}_{\mathcal{P}^{(k)}}^{(k)}$, and therefore outputs a single scalar.
Finally, $l^{(k)}$ is fed to $\text{CNN}_{\Em}$ as the matrix $l^{(k)} \cdot \mymathbb{1}_{N_f \times N_t}$.
The architecture of all \glspl{CNN} are detailed in Section~\ref{sec:architectures}.

\subsection{ML-enhanced demapper}
\label{sec:ml_demapper}

The sub-optimal channel estimation and equalization create distortions in the equalized signals.
%
As seen in Section~\ref{sec:conventional_architecture}, a conventional demapper processes each user and \gls{RE} independently.
In this work, we take a different approach and propose a new \gls{CNN}-based demapper, denoted by $\text{CNN}_{\text{Dmp}}$, that operates directly on the two-dimensional \gls{RG}. 
The demapping is still applied to each user independently (see Fig.\ref{fig:ml_receiver}), preserving the scalability of the conventional architecture, but the joint processing of all \glspl{RE} allows $\text{CNN}_{\text{Dmp}}$ to estimate and correct the distortions present in the equalized signals. 
$\text{CNN}_{\text{Dmp}}$ takes six inputs, each of size $N_f \times N_t$.
Similarly to $\text{CNN}_{\Em}$, the first two inputs are the positioning matrices, and the third input is the \gls{SNR}.
The forth input contains the post-equalization noise variances $\rho^2_{f,t,k}$ for all \glspl{RE} $(f,t)$.
The fifth and sixth inputs are the real and imaginary parts of the equalized symbols $\widehat{\Xm}^{(k)}$.
The output of $\text{CNN}_{\text{Dmp}}$ is of dimension $N_f \times N_t \times M $, corresponding to the \glspl{LLR} of user $k$.

\begin{table}[t]
\vspace{2pt}
  \centering
  \renewcommand{\arraystretch}{1.25}
  \begin{tabular}{|p{1.8cm}||c|c|c||c|c|c|}
    \hline
     & \multicolumn{3}{c||}{$\text{CNN}_{l}$} & \multicolumn{3}{c|}{$\text{CNN}_{\text{Dmp}}$} \\ \hhline{|=||===||===|}
    

    Parameters  & filters & kernel & dilat.  & filters & kernel & dilat.  \\

    \hline
    Conv2D        & 32 & (1,1) & (1,1) & 128 & (1,1) & (1,1)   \\ \hline
    ResNet Layer  & 32 & (3,2) & (1,1) & 128 & (3,3) & (1,1)   \\ \hline
    ResNet Layer  & 32 & (5,2) & (2,1) & 128 & (5,3) & (2,1)   \\ \hline
    ResNet Layer  & 32 & (7,2) & (3,1) & 128 & (7,3) & (3,2)   \\ \hline
    ResNet Layer  & 32 & (5,2) & (2,1) & 128 & (9,3) & (4,3)   \\ \hline
    ResNet Layer  & 32 & (3,2) & (1,1) & 128 & (7,3) & (3,2)   \\ \hline
    ResNet Layer  & \multicolumn{3}{c||}{-} & 128 & (5,3) & (2,1)    \\ \hline
    ResNet Layer  & \multicolumn{3}{c||}{-} & 128 & (3,3) & (1,1)   \\ \hline
    Conv2D  		 & 1 &(3,2) & (1,1) & $M$  & (1,1) & (1,1)  \\ \hline
    
    Output Layer & \multicolumn{3}{c||}{Dense, units = 1}  & \multicolumn{3}{c|}{-} \\  \hline
    
    
  \end{tabular}
  
  \caption{Architectures of the different CNNs.}
  \label{table:CNNs}
  \vspace{-10pt}
\end{table}

\subsection{CNN architectures}
\label{sec:architectures}

$\text{CNN}_{\Em}$, $\text{CNN}_{l}$, and $\text{CNN}_{\text{Dmp}}$ share similar architectures.
$\text{CNN}_{\Em}$ uses two convolutional 2D layers with 32 filters, kernel sizes of $(5,3)$, dilation rates of $(1,1)$, zero-padding, and \gls{ReLU} activation functions.
Its output layer is a convolutional 2D layer but with only two filters, a kernel size and dilation rate of $(1,1)$, zero-padding, and sigmoid activation function.
Both $\text{CNN}_{l}$ and $\text{CNN}_{\text{Dmp}}$ use custom \emph{ResNet} layers, inspired by \cite{he2016identity}, and composed of a batch normalization layer, a \gls{ReLU}, a separable convolutional 2D layer, and followed by the addition of the input, as shown in Fig.~\ref{fig:resnet}.
Separable convolutional 2D layers are known to be more efficient than traditional convolutional layers \cite{howard2017mobilenets}.
The architecture of  $\text{CNN}_{l}$ and $\text{CNN}_{\text{Dmp}}$ are detailed in Table~\ref{table:CNNs}, where all convolutional layers use zero-padding to keep consistent dimensions.
Varying kernel sizes and dilation rates were used to increase the receptive field of the \glspl{CNN} \cite{korpi2020deeprx}.

\begin{figure}[t]
\centering
\includegraphics[width=0.45\textwidth]{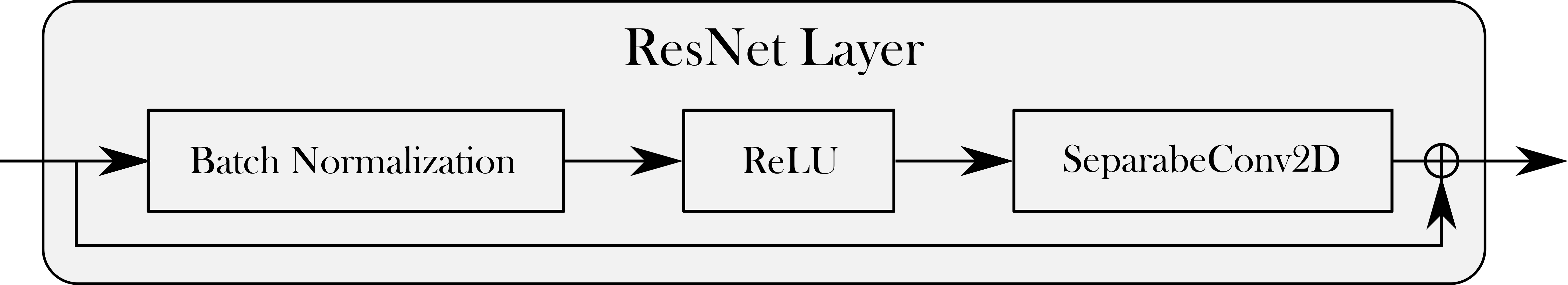}
  \captionof{figure}{Custom ResNet layer.}
  \label{fig:resnet}
    \vspace{-15pt}
\end{figure}

\section{Evaluations} 
\label{sec:evaluations}

\begin{figure*}

\centering
	\begin{subfigure}[c]{1\textwidth}
	\centering
	\begin{tikzpicture} 
	
	\definecolor{color0}{rgb}{0.12156862745098,0.466666666666667,0.705882352941177}
	\definecolor{color1}{rgb}{1,0.498039215686275,0.0549019607843137}
	\definecolor{color2}{rgb}{0.172549019607843,0.627450980392157,0.172549019607843}
	\definecolor{color3}{rgb}{0.83921568627451,0.152941176470588,0.156862745098039}

    	\begin{axis}[%
    	hide axis,
    	xmin=10,
   	 xmax=50,
    	ymin=0,
    	ymax=0.4,
    	legend columns=4, 
    	legend style={draw=white!15!black,legend cell align=left,column sep=6.95ex}
    	]
    
    	\addlegendimage{white,mark=*, mark size=2}
    	\addlegendentry{\hspace{-1 cm} Spectral interp. :};
    	\addlegendimage{color0,mark=*, mark size=2}
    	\addlegendentry{\hspace{-0.5cm}Baseline};
    	\addlegendimage{color2, mark=diamond*, mark size=2}
    	\addlegendentry{\hspace{-0.5cm}ML receiver};
    	\addlegendimage{color3, mark=pentagon*, mark size=2}
    	\addlegendentry{\hspace{-0.5cm} Perfect CSI \quad \quad \hspace{1cm}};
    	\end{axis}
	\end{tikzpicture}
	\end{subfigure}%
	\vspace{3pt}

  	\begin{subfigure}{0.4\textwidth}
  	
	\begin{tikzpicture} [trim left=3.57cm]
	
	\definecolor{color2}{rgb}{0.172549019607843,0.627450980392157,0.172549019607843}

    	\begin{axis}[%
    	hide axis,
    	xmin=10,
   	xmax=50,
    	ymin=0,
    	ymax=0.4,
    	legend columns=2, 
    	legend style={draw=white!15!black,legend cell align=right,column sep=0.5ex}
    	]
    	\addlegendimage{white,mark=square*, mark size=2}
    	\addlegendentry{\hspace{-0.9cm} 1P pattern :};
    	\addlegendimage{color2, dotted, mark=x, mark size=2, mark options={solid}}
    	\addlegendentry{\hspace{-0.05cm} ML receiver trained with $N_k=2$};
    	\end{axis}
	\end{tikzpicture}
	\vspace{0cm}
	\end{subfigure}
	

  	\begin{subfigure}{0.4\textwidth}
  	
	\vspace{-16.7pt}
	\begin{tikzpicture} [trim left=-5.2cm]
	
	\definecolor{color0}{rgb}{0.12156862745098,0.466666666666667,0.705882352941177}
	\definecolor{color2}{rgb}{0.172549019607843,0.627450980392157,0.172549019607843}

    	\begin{axis}[%
    	hide axis,
    	xmin=10,
   	 xmax=50,
    	ymin=0,
    	ymax=0.4,
    	legend columns=3, 
    	legend style={draw=white!15!black,legend cell align=left, column sep=0.5ex}
    	]
    	\addlegendimage{white,mark=square*, mark size=2}
    	\addlegendentry{\hspace{-0.9cm} 2P\,pattern,\,dual\,interp.\,:};
    	\addlegendimage{color0, dashed, mark=*, mark size=1, mark options={solid}}
    	\addlegendentry{\hspace{-0.15cm}  Baseline};
    	\addlegendimage{color2, dashed, mark=+, mark size=2, mark options={solid}}
    	\addlegendentry{\hspace{-0.15cm}  ML rec.};
    	\end{axis}
	\end{tikzpicture}
	\vspace{5pt}	
	\end{subfigure}%

  \centering
  	\begin{subfigure}[b]{0.27\textwidth}
  	\vspace{-10pt}
    	\begin{adjustbox}{width=\linewidth} 
\begin{tikzpicture}

\definecolor{color0}{rgb}{0.12156862745098,0.466666666666667,0.705882352941177}
\definecolor{color1}{rgb}{1,0.498039215686275,0.0549019607843137}
\definecolor{color2}{rgb}{0.172549019607843,0.627450980392157,0.172549019607843}
\definecolor{color3}{rgb}{0.83921568627451,0.152941176470588,0.156862745098039}

\begin{axis}[
log basis y={10},
tick align=outside,
tick pos=left,
x grid style={white!69.0196078431373!black},
xlabel={SNR},
xmajorgrids,
xmin=-5.75, xmax=10.75,
xtick style={color=black},
y grid style={white!69.0196078431373!black},
ylabel={BER},
ymajorgrids,
ymin=9e-06, ymax=0.284996815115009,
ymode=log,
ytick style={color=black}
]
\addplot [semithick, color0, mark=*, mark size=2, mark options={solid}]
table {%
-5 1.70455799e-01
-2.5 7.13982590e-02
0 1.61042529e-02
2.5 3.68813382e-03
5 8.11323304e-04
7.5 1.83531746e-04
10 3.74090606e-05
};

\addplot [semithick, color2, mark=diamond*, mark size=2, mark options={solid}]
table {%
-5 1.63974658e-01
-2.5 6.29884541e-02 
0.0 1.35914254e-02 
2.5 2.97567378e-03
5.0 5.10361550e-04 
7.5 1.17256394e-04 
10.0 1.54320987e-05
};

\addplot [semithick, dotted, color2, mark=x, mark size=2, mark options={solid}]
table {%
-5 0.16636697947978973
-2.5 0.06406939029693604
0 0.014092790834125011
2.5 0.0031420166441239418
5 0.0005471505739842542
7.5 0.00013356316078396048
10 2.0864197105984206e-05
};
 
\addplot [semithick, color3, mark=pentagon*, mark size=2, mark options={solid}]
table {%
-5 9.92080718e-02 
-2.5 2.60168653e-02 
0 3.39599513e-03 
2.5 5.64787257e-04
5 8.90790339e-05 
7.5 1.18736220e-05 
10 7.03339944e-07
};
\end{axis}

\end{tikzpicture}
  	\end{adjustbox}  
  	\vspace{-17pt}
  	\caption{1P pilot pattern at \SIrange[range-units=single]{0}{15}{\km\per\hour}.}
  	\label{fig:UL_1P_L}
	\end{subfigure}%
  	\hfill
  	\begin{subfigure}[b]{0.27\textwidth}
  	\vspace{-10pt}
    	\begin{adjustbox}{width=\linewidth} 
\begin{tikzpicture}

\definecolor{color0}{rgb}{0.12156862745098,0.466666666666667,0.705882352941177}
\definecolor{color1}{rgb}{1,0.498039215686275,0.0549019607843137}
\definecolor{color2}{rgb}{0.172549019607843,0.627450980392157,0.172549019607843}
\definecolor{color3}{rgb}{0.83921568627451,0.152941176470588,0.156862745098039}

\begin{axis}[
log basis y={10},
tick align=outside,
tick pos=left,
x grid style={white!69.0196078431373!black},
xlabel={SNR},
xmajorgrids,
xmin=-5.75, xmax=10.75,
xtick style={color=black},
y grid style={white!69.0196078431373!black},
ylabel={BER},
ymajorgrids,
ymin=4.45495867691486e-05, ymax=0.290400883476259,
ymode=log,
ytick style={color=black}
]
\addplot [semithick, color0, mark=*, mark size=2, mark options={solid}]
table {%
-5 0.1846478134393692
-2.5 0.08860826243956883
0 0.02830456556486232
2.5 0.008724582690774696
5 0.003121509410751363
7.5 0.0011042181134020212
10 0.0004941023669240529
};

\addplot [semithick, color2, mark=diamond*, mark size=2, mark options={solid}]
table {%
-5 0.1726982742547989
-2.5 0.07286844154198964
0 0.02001202671921679
2.5 0.00480035580767435
5 0.00123374117470424
7.5 0.00027827050336578395
10 8.65634173876606e-05
};

\addplot [semithick, dotted, color2, mark=x, mark size=2, mark options={solid}]
table {%
-5 0.1749269664287567
-2.5 0.07480434576670329
0 0.020850310434720347
2.5 0.005127921929670265
5 0.0013553608214715495
7.5 0.0003655563643202963
10 0.00011573243706958602
};

\addplot [semithick, color3, mark=pentagon*, mark size=2, mark options={solid}]
table {%
-5 0.11705315858125687
-2.5 0.037146118779977165
0 0.00776083000736045
2.5 0.001644020157527848
5 0.00036343511077575386
7.5 7.203740497866723e-05
10 2.1999927072708183e-05
};
\end{axis}

\end{tikzpicture}
  	\end{adjustbox}  
  	\vspace{-17pt}
  	\caption{1P pilot pattern at \SIrange[range-units=single]{15}{30}{\km\per\hour}.}  %
  	\label{fig:UL_1P_M}
	\end{subfigure}%
   \hfill
   \begin{subfigure}[b]{0.27\textwidth}
   \vspace{-10pt}
    	\begin{adjustbox}{width=\linewidth} 
\begin{tikzpicture}

\definecolor{color0}{rgb}{0.12156862745098,0.466666666666667,0.705882352941177}
\definecolor{color1}{rgb}{1,0.498039215686275,0.0549019607843137}
\definecolor{color2}{rgb}{0.172549019607843,0.627450980392157,0.172549019607843}
\definecolor{color3}{rgb}{0.83921568627451,0.152941176470588,0.156862745098039}

\begin{axis}[
log basis y={10},
tick align=outside,
tick pos=left,
x grid style={white!69.0196078431373!black},
xlabel={SNR},
xmajorgrids,
xmin=-5.75, xmax=10.75,
xtick style={color=black},
y grid style={white!69.0196078431373!black},
ylabel={BER},
ymajorgrids,
ymin=0.000204329461932803, ymax=0.296876644347008,
ymode=log,
ytick style={color=black}
]
\addplot [semithick, color0, mark=*, mark size=2, mark options={solid}]
table {%
-5 0.1975618600845337
-2.5 0.10643325746059418
0 0.06169325932860374
2.5 0.029828645288944244
5 0.01696120185910591
7.5 0.010197379318997264
10 0.008980654748156666
};

\addplot [semithick, color2, mark=diamond*, mark size=2, mark options={solid}]
table {%
-5 0.18415522575378418
-2.5 0.07863308861851692
0 0.031500082835555075
2.5 0.009625039558159187
5 0.003806682806628357
7.5 0.001328262786992127
10 0.0005091903651191388
};

\addplot [semithick, dotted, color2, mark=x, mark size=2, mark options={solid}]
table {%
-5 0.1855296492576599
-2.5 0.08241016417741776
0 0.03380377014875412
2.5 0.010749311887947845
5 0.004333014971197449
7.5 0.00171167318113148
10 0.0007236755946837365
};

\addplot [semithick, color3, mark=pentagon*, mark size=2, mark options={solid}]
table {%
-5 0.15017016232013702
-2.5 0.057083263993263245
0 0.022171241417527198
2.5 0.006593519210582599
5 0.002502959137061788
7.5 0.0007695381435769377
10 0.00039520943708950654
};
\end{axis}

\end{tikzpicture}
   		 
  	\end{adjustbox}    
  	\vspace{-17pt}
  	\caption{1P pilot pattern at \SIrange[range-units=single]{30}{45}{\km\per\hour}.}   %
  	\label{fig:UL_1P_H}
	\end{subfigure}%
   
   \vspace{7pt}
  	
  	\begin{subfigure}[b]{0.27\textwidth}
  	\begin{adjustbox}{width=\linewidth} 
\begin{tikzpicture}

\definecolor{color0}{rgb}{0.12156862745098,0.466666666666667,0.705882352941177}
\definecolor{color1}{rgb}{1,0.498039215686275,0.0549019607843137}
\definecolor{color2}{rgb}{0.172549019607843,0.627450980392157,0.172549019607843}
\definecolor{color3}{rgb}{0.83921568627451,0.152941176470588,0.156862745098039}

\begin{axis}[
log basis y={10},
tick align=outside,
tick pos=left,
x grid style={white!69.0196078431373!black},
xlabel={SNR},
xmajorgrids,
xmin=-5.75, xmax=10.75,
xtick style={color=black},
y grid style={white!69.0196078431373!black},
ylabel={BER},
ymajorgrids,
ymin=1.92917347126291e-05, ymax=0.296394799327145,
ymode=log,
ytick style={color=black}
]
\addplot [semithick, color0, mark=*, mark size=2, mark options={solid}]
table {%
-5 0.17254532873630524
-2.5 0.07584972977638245
0 0.02174021042883396
2.5 0.005923466421081685
5 0.0016146315616788344
7.5 0.0005304822564478674
10 0.00032807677547680214
};

\addplot [semithick, color2, mark=diamond*, mark size=2, mark options={solid}]
table {%
-5 0.1668788567185402
-2.5 0.06520061716437339
0 0.01731626146938652
2.5 0.004077305170940235
5 0.0009451195951260161
7.5 0.00023105517006712033
10 9.108410478394945e-05
};

\addplot [semithick, color3, mark=pentagon*, mark size=2, mark options={solid}]
table {%
-5 0.11548755690455437
-2.5 0.034257329627871515
0 0.007366174703929573
2.5 0.0013924575551209272
5 0.000289592979097506
7.5 7.14216814685642e-05
10 3.477044709143229e-05
};

\addplot [semithick, dashed, color0, mark=*, mark size=1, mark options={solid}]
table {%
-5 0.16387683898210526
-2.5 0.07110339589416981
0 0.0158540703356266
2.5 0.004082609954057261
5 0.0009156539328250801
7.5 0.00025631751583205187
10 0.00010894097156779026
};

\addplot [semithick, dashed, color2, mark=+, mark size=2, mark options={solid}]
table {%
-5 0.16032986342906952
-2.5 0.06646010279655457
0 0.014224054881681998
2.5 0.003549913188617211
5 0.0007263696057179913
7.5 0.0001927083330519963
10 8.029513934161514e-05
};

\end{axis}

\end{tikzpicture}
  	\end{adjustbox} 
  	\vspace{-17pt}
  	\caption{2P pilot pattern at \SIrange[range-units=single]{50}{70}{\km\per\hour}.}
  	\label{fig:UL_2P_L}
	\end{subfigure}%
	\hfill
  	\begin{subfigure}[b]{0.27\textwidth}
  	\begin{adjustbox}{width=\linewidth} 
\begin{tikzpicture}

\definecolor{color0}{rgb}{0.12156862745098,0.466666666666667,0.705882352941177}
\definecolor{color1}{rgb}{1,0.498039215686275,0.0549019607843137}
\definecolor{color2}{rgb}{0.172549019607843,0.627450980392157,0.172549019607843}
\definecolor{color3}{rgb}{0.83921568627451,0.152941176470588,0.156862745098039}

\begin{axis}[
log basis y={10},
tick align=outside,
tick pos=left,
x grid style={white!69.0196078431373!black},
xlabel={SNR},
xmajorgrids,
xmin=-5.75, xmax=10.75,
xtick style={color=black},
y grid style={white!69.0196078431373!black},
ylabel={BER},
ymajorgrids,
ymin=3e-5, ymax=0.29056388782237,
ymode=log,
ytick style={color=black}
]
\addplot [semithick, color0, mark=*, mark size=2, mark options={solid}]
table {%
-5 0.19766107201576233
-2.5 0.09440104166666667
0 0.04122154731303453
2.5 0.01689288727092472
5 0.006960069447134932
7.5 0.003392811199494948
10 0.002067515436404695
};

\addplot [semithick, color2, mark=pentagon*, mark size=2, mark options={solid}]
table {%
-5 0.18930844962596893
-2.5 0.07495659838120143
0 0.025046296417713165
2.5 0.007426112948451191
5 0.0018183513368542966
7.5 0.0005286265418462184
10 0.00013654578315249334
};

\addplot [semithick, color3, mark=diamond*, mark size=2, mark options={solid}]
table {%
-5 0.15181809663772583
-2.5 0.05091467499732971
0 0.015293209999799728
2.5 0.0044383358979371915
5 0.0010596707855196048
7.5 0.0002891589513455983
10 9.098508347960887e-05
};

\addplot [semithick, dashed, color0, mark=*, mark size=1, mark options={solid}]
table {%
-5 0.1812524050474167
-2.5 0.08607855997979641
0 0.025863500725891855
2.5 0.006777777799094717
5 0.0016710648123097296
7.5 0.00044164094332809324
10 0.00010615997982919605
};

\addplot [semithick, dashed, color2, mark=+, mark size=2, mark options={solid}]
table {%
-5 0.17500963807106018
-2.5 0.07563898526132107
0 0.021413376710067194
2.5 0.004997427999041975
5 0.000994997425202746
7.5 0.0002412808630935615
10 5.542695515032392e-05
};

\end{axis}
\end{tikzpicture}  		
  	\end{adjustbox} 
  	\vspace{-17Pt}
  	\caption{2P pilot pattern at \SIrange[range-units=single]{80}{100}{\km\per\hour}.}
  	\label{fig:UL_2P_M}
	\end{subfigure}%
    \hfill
  	\begin{subfigure}[b]{0.27\textwidth}
  	\begin{adjustbox}{width=\linewidth} 
\begin{tikzpicture}

\definecolor{color0}{rgb}{0.12156862745098,0.466666666666667,0.705882352941177}
\definecolor{color1}{rgb}{1,0.498039215686275,0.0549019607843137}
\definecolor{color2}{rgb}{0.172549019607843,0.627450980392157,0.172549019607843}
\definecolor{color3}{rgb}{0.83921568627451,0.152941176470588,0.156862745098039}

\begin{axis}[
log basis y={10},
tick align=outside,
tick pos=left,
x grid style={white!69.0196078431373!black},
xlabel={SNR},
xmajorgrids,
xmin=-5.75, xmax=10.75,
xtick style={color=black},
y grid style={white!69.0196078431373!black},
ylabel={BER},
ymajorgrids,
ymin=0.000254659159228147, ymax=0.289449661539661,
ymode=log,
ytick style={color=black}
]
\addplot [semithick, color0, mark=*, mark size=2, mark options={solid}]
table {%
-5 0.21111111342906952
-2.5 0.13478250056505203
0 0.07800082117319107
2.5 0.0415351428091526
5 0.025429012477397917
7.5 0.01752727863419315
10 0.015883873384445905
};

\addplot [semithick, color2, mark=pentagon*, mark size=2, mark options={solid}]
table {%
-5 0.20206403732299805
-2.5 0.10108266025781631
0 0.03466493115574121
2.5 0.010000613729723475
5 0.0036591435558511877
7.5 0.001552121073156899
10 0.0007995756181480829
};
\addplot [semithick, color3, mark=diamond*, mark size=2, mark options={solid}]
table {%
-5 0.17277199029922485
-2.5 0.09070698171854019
0 0.038450183674693106
2.5 0.014035318343138153
5 0.0059697145223617555
7.5 0.0032043620683354043
10 0.0021673418187128845
};

\addplot [semithick, dashed, color0, mark=*, mark size=1, mark options={solid}]
table {%
-5 0.2037760466337204
-2.5 0.11408339689175288
0 0.041163676977157594
2.5 0.014065962368383622
5 0.005237557886168361
7.5 0.0019424768530327129
10 0.0008357445964429644
};

\addplot [semithick, dashed,  color2, mark=+, mark size=2, mark options={solid}]
table {%
-5 0.19689911603927612
-2.5 0.0950665498773257
0 0.030419078283011915
2.5 0.008801489799784927
5 0.002578800140647218
7.5 0.0007801118886767654
10 0.00034008487333267115
};

\end{axis}

\end{tikzpicture}
  	\end{adjustbox} 
  	\vspace{-17pt}
  	\caption{2P pilot pattern at \SIrange[range-units=single]{110}{130}{\km\per\hour}.}
  	\label{fig:UL_2P_H}
	\end{subfigure}%

\vspace{5pt}
\caption{Uplink BER achieved by the different receivers with the 1P and 2P pilot patterns.}
\vspace{-10pt}
\label{fig:eval_UL}
\end{figure*}
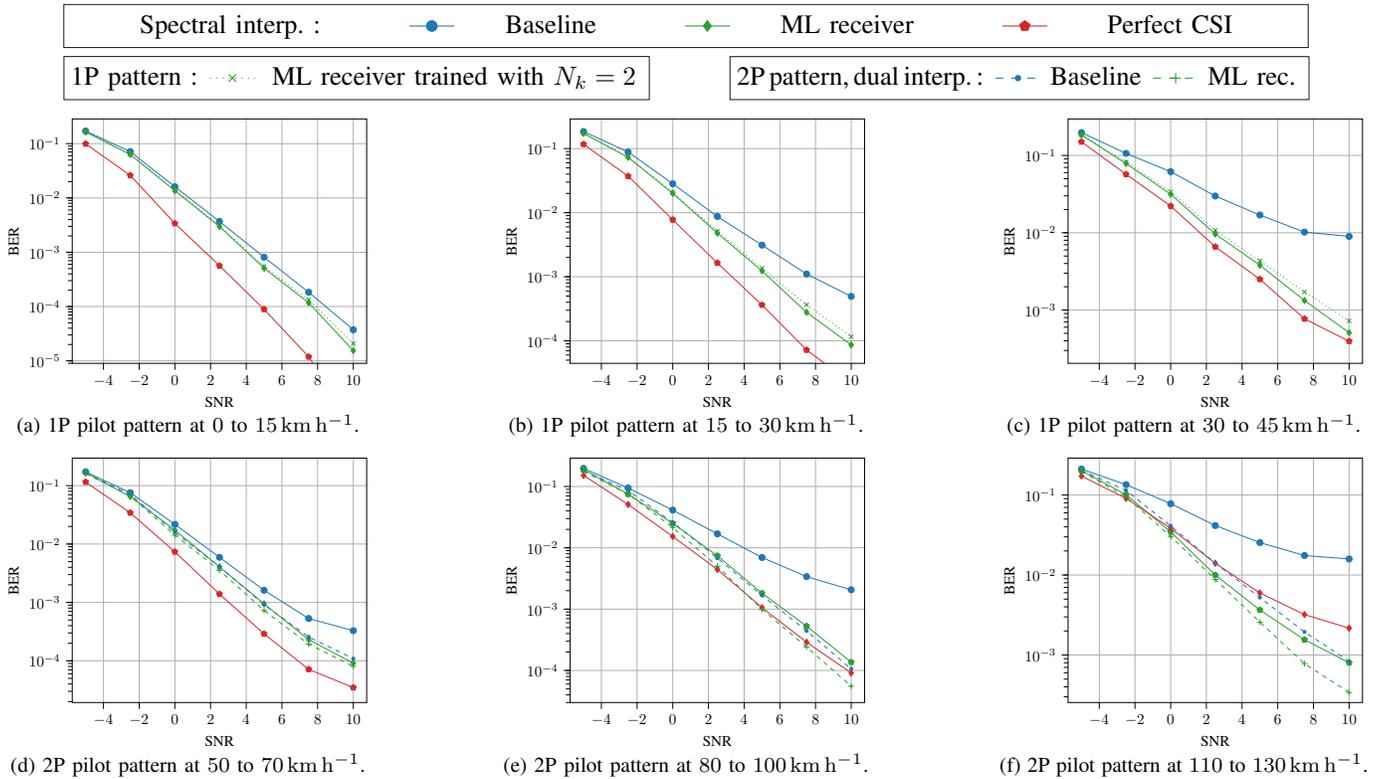


\subsection{Training and evaluation setup}

The number of users and receive antennas were set to $N_k =4$ and $N_m = 16$, respectively.
Each user was randomly positioned between \SI{15}{} and \SI{150}{\m} of the \gls{BS} in a \SI{120}{\degree} cell sector.
The users and \gls{BS} heights were respectively set to \SI{1.5}{\m} and \SI{10}{\m}.
Realistic channel realizations were generated with QuaDRiGa version 2.0.0 using the 3GPP \gls{NLOS} \gls{UMi} channel model.
The \glspl{RG} were composed of $N_t = 14$ \gls{OFDM} symbols and $N_f = 72$ subcarriers (six resource blocks).
The center frequency was set to \SI{3.5}{\GHz} and the subcarrier spacing to \SI{15}{\kHz}.
A Gray-labeled $16$-QAM was used to transmit $M=4$ bits per symbol.
Six ranges of user speed were considered: \SIrange[range-units=single]{0}{15}{\km\per\hour}, \SIrange[range-units=single]{15}{30}{\km\per\hour}, and \SIrange[range-units=single]{30}{45}{\km\per\hour} for the 1P pilot pattern, and  \SIrange[range-units=single]{50}{70}{\km\per\hour}, \SIrange[range-units=single]{80}{100}{\km\per\hour}, and  \SIrange[range-units=single]{110}{130}{\km\per\hour} for the 2P pilot pattern.
We noticed that $\text{CNN}_l$ was not able to extract useful information from the 1P pattern, and therefore was not used in the corresponding training and evaluations.

Two separate sets were considered to train the \gls{ML}-enhanced receiver on the 1P and 2P pilot patterns.
Both training sets were made of 3000 \glspl{RG} constructed from 1000 \glspl{RG} of the three users speed ranges for each pilot pattern.
The receiver parameters were randomly initialized, except for $\gamma$ that was initialized at $\pi$.
The Adam optimizer was used for training, with a batch size of $B_s = 27$ and a learning rate of $10^{-3}$.
During evaluations, a standard IEEE 802.11n \gls{LDPC} code of length \SI{1296}{\bit} was used in conjunction with 40 iterations of a conventional belief-propagation decoder.
To satisfy the perfect power allocation assumption, the average energy per user and antenna on the \glspl{RG} were normalized such that $\sum_{f=1}^{N_f} \sum_{t=1}^{N_t} ||\hv_{f,t}^{(k)}||_2^2 = N_f N_t N_m$.

\subsection{Simulation results}

The proposed \gls{ML}-enhanced receiver was compared against two systems.
The first one is the conventional receiver baseline detailed in Section~\ref{sec:conventional_architecture}.
The second one, referred to as `Perfect CSI receiver', has the same architecture but has access to perfect knowledge of the channel at \glspl{RE} carrying pilots and to $\Em_{f,t}$ at every \gls{RE}, obtained from Monte-Carlo simulations.
Additional simulations were conducted for the 2P pilot pattern using spectral and temporal interpolation for both the baseline and the \gls{ML} receiver.
Finally, an ML receiver trained with only $N_k = 2$ users was also evaluated with $N_k = 4$ users.
The first row of Fig.~\ref{fig:eval_UL} presents evaluation results corresponding to the 1P pattern.
It can be seen that at low speed (\SIrange[range-units=single]{0}{15}{\km\per\hour}), the \gls{ML}-enhanced receiver achieves a \SI{0.8}{\dB} gain over the conventional receiver at a coded \gls{BER} of $10^{-3}$, but is still \SI{2.3}{\dB} behind the perfect \gls{CSI} receiver.
Between $35$ and \SI{45}{\km\per\hour}, the conventional receiver saturates at high \gls{SNR} while the gap between the \gls{ML}-enhanced and perfect \gls{CSI} receivers is narrowing, leading to a gain of \SI{2.5}{\dB} at a \gls{BER} of $10^{-2}$.
The \gls{ML} receiver trained with a mismatched number of users only suffers a small performance drop, demonstrating the scalability of the \gls{ML} scheme.
The 2P pattern results shown in the second row of Fig.~\ref{fig:eval_UL} follow the same trend, with increasing gains at higher speeds.
In the \SIrange[range-units=single]{110}{130}{\km\per\hour} range, the \gls{ML}-enhanced receiver even outperforms the perfect \gls{CSI} baseline, which suffers from strong channel aging that the \gls{CNN} demapper is able to mitigate.
Using both spectral and temporal interpolations reduces the gains provided by the \gls{ML} receiver, which can be explained by the better channel estimates leading to less channel aging, but they still amount to a \SI{2.2}{\dB} gap at a \gls{BER} of $10^{-3}$ for the highest speeds.
More detailed uplink and downlink evaluation results can be found in \cite{goutay2020machine}, including a more detailed analysis on the interpretability and usefulness of each \gls{ML} component.

\section{Conclusion}
\label{sec:conclusion}

In this paper, we have enhanced a conventional \gls{MU}-\gls{MIMO} receiver architecture using multiple \glspl{CNN} to improve both the computation of its channel estimation error statistics and of the estimated bit probabilities.
Compared to other approaches, all \gls{ML} components of the receiver are jointly optimized to maximize the information rate of the transmission, which does not require any ground truth measurements.
Evaluation results indicate that the \gls{ML} receiver achieves gains across all scenarios, and especially at high speeds, while preserving the scalability and interpretability of conventional architectures.


\bibliographystyle{IEEEtran}
\bibliography{IEEEabrv, bib_abrv, bibliography}

\end{NoHyper}
\end{document}